\documentclass[prl,twocolumn,showpacs,aps,floatfix,amsmath,amssymb,amsfonts]{revtex4}
\usepackage{graphicx}
\usepackage{subfigure}

\newcommand{\mrm}{\mathrm}

\begin{document}

\title{Quantum theory of reactive collisions for $1/r^n$ potentials}
\author{Krzysztof Jachymski$^{1}$, Micha{\l} Krych$^{1}$, Paul S. Julienne$^{2}$, and Zbigniew Idziaszek$^{1}$}
\affiliation{$^1$Faculty of Physics, University of Warsaw, Ho{\.z}a 69,
00-681 Warsaw, Poland,\\
$^{2}$Joint Quantum Institute, NIST and
the University of Maryland, Gaithersburg, Maryland 20899-8423, USA}


\date{\today}

\begin{abstract}
We develop a general quantum theory for reactive collisions involving power-law potentials ($-1/r^n$) valid from the ultracold up to the high-temperature limit. Our quantum defect framework extends the conventional capture models to include the non-universal case when the short-range reaction probability $P^\mathrm{re}<1$. We present explicit analytical formulas as well as numerical studies for the van der Waals ($n=6$) and polarization ($n=4$) potentials. Our model agrees well with recent merged beam experiments on Penning ionization, spanning collision energies from $10$mK to $30$K [Henson et al, Science 338, 234(2012)].
\end{abstract}

\maketitle

Much recent work involves the inelastic and reactive collisions of cold atoms or molecules with one another~\cite{Hudson2008,Ospelkaus2010,Miranda2011,Deiglmayr2011a} or with ions in hybrid traps~\cite{Hudson2011,Kohl2012,Willitsch2012}. These could involve the ultracold regime with translational temperature on the order of $\mu$K or less or the cold regime between a few mK and a few K. Systematic theoretical principles for understanding the quantum dynamics of such collisions are needed.  Much work has already been done in this area, as reviewed by Ref.~\cite{PSJ2012}. One class of theories based on Quantum Defect Theory (QDT) allows us to systematize and develop tools for understanding such collisions~\cite{Fano,Mies1984a,Mies1984b,Julienne1989,Burke1998,Gao2008,Nikitkin}. One special limiting case is that of highly reactive collisions, where simple  classical trajectory capture models known as the Langevin ($n=4$)~\cite{Langevin1905} or Gorin ($n=6$)~\cite{Gorin1938} models apply when the long range potential takes on the form $-C_n/r^n$ ($n>3$). These familiar models assume that every classical trajectory contributes to the collision cross section that is captured by the long range potential so the particles spiral in to short distance where they react or relax with probability $P^\mathrm{re}$. We use the term "universal" to describe capture models with $P^{\mathrm{re}}=1$, since they do not depend on any details of the strong short-range chemical interactions.

In the cold and ultracold regimes, it is essential to build in quantum corrections to these classical models due to quantum threshold laws~\cite{Orzel1999,Quemener2010,Idziaszek2010R}. This has been done using QDT for both the Langevin~\cite{Gao2011,NJP2011} and Gorin~\cite{Idziaszek2010,Gao2010PRL} universal models where $P^\mathrm{re} = 1$. Here, we will follow the formalism by Idziaszek {\it et al.}~\cite{Idziaszek2010,NJP2011} and generalize the previous results to the nonuniversal regime with $P^\mathrm{re} < 1$, and to the arbitrary collision energy.
In the limiting cases of low and high temperatures we give analytic formulas that are valid for power-law potentials $(-1/r^n)$. We apply our theory to
interpret the ionization rate constants measured by recent merged beam experiments in the cold regime~\cite{Narevicius}. Using a single complex QDT parameter found by fitting low energy data only, we are able to reproduce the experimental data over four orders of magnitude in energy including about twenty partial waves in the calculation.

{\it Generalized complex scattering length.}
We consider reactive collisions of particles interacting via power-law potential $V(r)=-C_n/r^n$ ($n>3$) at large distances $r \gtrsim R_0$, where $R_0$ denotes the range of short-range forces. These include, for example, collisions of $S$-state atoms ($n=6$), or collisions of $S$-state atoms with ions ($n=4$) \cite{Seaton1977,Drachman1979,Idziaszek2009a}. The characteristic length $R_n = \left(2 \mu C_n/\hbar^2 \right)^{1/(n-2)}$ and energy $E_n = \hbar^2/(2 \mu R_n^2)$ are associated with the long-range potential. Typically, $E_n$ is of the order of the centrifugal barrier height for $p$-waves, while $R_n$ defines the characteristic distance for the position of the barrier. The scattering channels are defined in terms of their internal states and the partial-wave quantum numbers $\ell\,m$. For simplicity we will use only $\ell\,m$ quantum numbers to specify the channels, while the internal quantum numbers will affect only the quantum-statistical prefactor which will be discussed later. Following \cite{Hutson2007,Idziaszek2010} we define an energy-dependent \cite{Blume2002,Bolda2002} complex scattering length for each $\ell$, $m$
\begin{align}
\label{alm}
\tilde{a}_{\ell m}(E) & = \tilde{\alpha}_{\ell m}(E) - i \tilde{\beta}_{\ell m}(E) =
\frac{1}{i k} \frac{1-S_{\ell m,\ell m}}{1+S_{\ell m,\ell m}}.
\end{align}
defined in terms of the diagonal elements of $S$ matrix. The elastic $\mathcal{K}^\mrm{el}$ and the reactive $\mathcal{K}^\mrm{re}$ rate constants can be expressed as \cite{Idziaszek2010}
\begin{align}
\mathcal{K}^\mrm{el}(E)=\sum_{\ell,m}{\mathcal{K}^\mrm{el}_{\ell m}(E)} & = g \frac{\pi \hbar}{\mu k} \sum_{\ell,m}{\left| 1 - S_{\ell m,\ell m}(E) \right|^2}\,,   \label{Kel} \\
\mathcal{K}^\mrm{re}(E)=\sum_{\ell,m}{\mathcal{K}^\mrm{re}_{\ell m}(E)} & = g \frac{\pi \hbar}{\mu k} \sum_{\ell,m}{\left(1- |S_{\ell m,\ell m}(E)|^2\right)}\,, \label{Kloss}
\end{align}
where $k^2 = 2 \mu E/\hbar^2$ with $E$ denoting the total energy, $\mu$ the reduced mass, and $g$ is equal to $1$, except the case when the particles are indistinguishable and are in identical internal states. In the latter case, $g=2$ and $\ell$ is restricted to even (odd) numbers for bosons (fermions). Comparison with experimental data requires averaging over actual distribution of energies in the particular experimental setup.

{\it Quantum Defect Theory.} We adopt the formulation of QDT by Mies~\cite{Mies1984a,Mies1984b} to find a general formula for $\tilde{a}_{\ell m}$. Mies uses different QDT functions than in Gao's similar treatment~\cite{Gao2008}, thus allowing additional insight into the dynamics. The basic idea of QDT is to separate short-range and long-range properties of the wave function. To this end one introduces a pair of linearly independent solutions $\hat{f}(r,E)$ and $\hat{g}(r,E)$ for each partial wave $\ell$ that have local WKB-like normalization at short distances, as done in~\cite{Idziaszek2010}. One then parametrizes the wave function at short range using $\hat{f}$ and $\hat{g}$ functions
\begin{equation}
\label{Psi}
\Psi(r,E)= A(E) \left[\hat{f}(r,E) + Y(E,\ell) \hat{g}(r,E) \right]
\end{equation}
Here, $A(E)$ is the amplitude, and $Y(E,\ell)$ is the quantum-defect parameter, that plays a key role in the quantum-defect method, parameterizing the short-range behavior of $\Psi(r)$. We consider a situation when short-range forces act at distances much smaller than $R_n$, so that we may neglect the energy and angular-momentum dependence of $Y(E,\ell)$~\cite{Gao2005x,NJP2011}. Then we can represent $Y$ in the form $Y = - i y$, where the constant real parameter $y$ is related to the probability of reaction \cite{Idziaszek2010}. Using a WKB-like representation of $\hat{f}$ and $\hat{g}$ yields
\begin{equation}
\label{WKBshort}
\Psi(r) \sim \frac{\exp\left[-i \int^r\!k(x)\mrm{d}x\right]}{\sqrt{k(r)}} - \left(\frac{1-y}{1+y}\right) \frac{\exp\left[i \int^r\!k(x)\mrm{d}x\right]}{\sqrt{k(r)}},
\end{equation}
where $k(r) = \sqrt{2 \mu \left(E - U_\ell(r)\right)}/\hbar$ is the local wave vector, $U_\ell(r) = V(r) + \hbar^2 \ell(\ell+1)/(2 \mu r^2)$. The first term represents the incident flux of  particles, and the second one the reflected flux. We notice that $y=1$ corresponds to unit reaction probability
(no outgoing flux) while for $y=0$ there are no losses (incident and reflected fluxes have equal amplitudes) \cite{Idziaszek2010}. The reaction probability is $P^\mrm{re} = 1 - \left[ (1-y)/(1+y)\right]^2= 4 y/(1+y)^2$.

In QDT one introduces a second pair of linearly independent solutions that are normalized at $r \to \infty$: $f(r,E,\ell)\cong\sin\left(kr-\ell\pi/2+\xi\right)/\sqrt{k}$ and $g(r,E,\ell)\cong\cos\left(kr-\ell\pi/2+\xi\right)/\sqrt{k}$, where $\xi(E,\ell)$ is the scattering phase shift of the potential $V(r)$. They are related to $\hat{f}$ and $\hat{g}$ by $f(r) = C^{-1}(E,\ell) \hat{f}(r)$ and $g(r) = C(E,\ell)[\hat{g}(r)+\tan \lambda(E,\ell) \hat{f}(r)]$, where
$C(E,\ell)$ and $\tan \lambda(E,\ell)$ are QDT functions with well-known analytical properties (see Refs.~\cite{Mies1984a,Mies1984b,Idziaszek2010} for details). Using general QDT formulas connecting $S$ with $Y(E,\ell)$ \cite{Mies1984a}, we derive an exact expression for the energy-dependent complex scattering length, which is valid for arbitrary energy and partial wave $\ell$
\begin{align}
\label{alm1}
\tilde{a}_{\ell m}(E)  =
- \frac{1}{k}\tan \left[
\xi(E,\ell) - \tan^{-1} \left( \frac{y C^{-2}(E,\ell)}{i + y \tan \lambda(E,\ell)}
\right)\right].
\end{align}
Substituting \eqref{alm1} into \eqref{Kloss} we can express the reaction rate directly in terms of QDT functions
\begin{equation}
\mathcal{K}^\mrm{re}_{\ell m}= g \frac{h}{2 \mu k} P^\mrm{re} \frac{ C^{-2}(E,\ell) (1+y)^2}{(1+y C^{-2}(E,\ell))^2 +y^2 \tan^2 \lambda (E,\ell)}.
\end{equation}
The $C$ and $\tan{\lambda}$ functions build in the effect of shape resonances due to quasi-bound states inside the long-range centrifugal barrier $U_\ell(r)$. References~\cite{Mies1984b,Julienne2006,NJP2011} give examples of these threshold functions.

It is interesting to investigate the low- and the high-energy limits of $\mathcal{K}^\mrm{re}$. At low energies in case of distinguishable-particle collisions or indistinguishable boson-boson collisions, the main contribution comes from $s$-wave scattering, while for identical fermions it comes from the $p$-wave. We introduce
\begin{equation}
\bar{a}=\frac{\pi(n-2)^{(n-4)/(n-2)}}{\Gamma^2\left(\frac{1}{n-2}\right)}R_n,
\end{equation}
which is the mean scattering length for $-1/r^n$ potential \cite{Gribakin1993}, and the characteristic $p$-wave volume
\begin{equation}
\bar{V}=\frac{\pi}{9}\frac{(n-2)^{(n-8)/(n-2)}}{\Gamma^2\left(\frac{3}{n-2}\right)}R_n ^3.
\end{equation}
For $n=6$ $\bar{a}\approx0.48 R_6$ and $\bar{V}\approx0.12 R_6^3$. A key result of this letter is the limit behavior of respective $s$ and $p$ reactive rate constants in the threshold regime, given by
\begin{align}
\label{Kl0}
\mathcal{K}^\mrm{re}_{00} & \stackrel{E\to0}{\longrightarrow}  2 g \frac{h}{\mu} \bar{a} y \frac{1+(s-\nu)^2}{1+y^2(s-\nu)^2}, \\
\label{Kl1}
\mathcal{K}^\mrm{re}_{1m} & \stackrel{E\to0}{\longrightarrow}  2 g \frac{h}{\mu } \bar{V} k^2 y \frac{1+\nu^2}{\nu^2} \frac{1+(s-\nu)^2}{y^2(s-\nu+\nu^{-1})^2+(s\nu^{-1}-2)^2},
\end{align}
where $\nu=\cot\frac{\pi}{n-2}$ and $s=a/\bar{a}$ with $a$ denoting the $s$-wave scattering length of the potential $V(r)$. One can show that for a fixed reaction parameter $y$, the maximal reaction rate is obtained for $s \to \pm \infty$ ($\ell = 0$) and for $s \to 2\cot \frac{\pi}{n-2}$ ($\ell = 1$), which correspond to the resonant case of a bound state crossing the threshold. In the universal regime ($y=1$) Eqs.~\eqref{Kl0}-\eqref{Kl1} reduce to $\mathcal{K}^\mrm{re}_{00} \stackrel{E\to0}{\longrightarrow} 2 g \frac{h}{\mu} \bar{a}$ and $\mathcal{K}^\mrm{re}_{1m} \stackrel{E\to0}{\longrightarrow}  2 g \frac{h}{\mu} \bar{V} k^2$, independent of $s$.

At high energies where many partial waves contribute we first derive an approximate expression corresponding to the classical limit of the scattering. We assume $C^{-2}(E,\ell) =1$ and $\tan \lambda(E,\ell) =0$ for partial waves at which the collision takes place above the barrier, while for collisions below the barrier we take $C^{-2}(E,\ell) =0$ \cite{Mies1984a}. This neglects the effects of the quantum tunneling and of the quantum reflection. In this approximation we obtain
\begin{equation}
\mathcal{K}^\mrm{re} \stackrel{E\to\infty}{\longrightarrow}  g \frac{h}{2 \mu k} P^\mrm{re} \ell_\mrm{max}(E)\left[1+\ell_\mrm{max}(E)\right]
\end{equation}
where $\ell_\mrm{max}(E)$ is the maximal angular momentum at which the top of the barrier is equal to the collision energy $E$. For a power-law potential $V(r) = - C_n/r^n$ this leads to
\begin{equation}
\mathcal{K}^\mrm{re} \stackrel{E\to\infty}{\longrightarrow}   g \frac{h}{2 \mu k} P^\mrm{re} \frac{n}{2} \left( \frac{E/E_n}{\frac{n}{2}-1}\right)^{(n-2)/n}.
\label{KLang}
\end{equation}
This result corresponds to the picture from classical physics where all trajectories that fall into the collision center contribute to the reaction rate with the probability $P^\mrm{re}$.


A more accurate treatment requires inclusion of the quantum effects, in particular when the collision energy is close to the top of the barrier. In our approach we approximate the top of the centrifugal barrier with a parabolic barrier $\frac12 \mu \omega_\ell^2 (r-r_0)^2$
with imaginary frequency $\hbar\omega_\ell= i \frac{\sqrt{2n-4}}{\left(n/2\right)^{2/(n-2)}} E_n \left[\ell(\ell+1)\right]^{\frac{n+2}{2n-4}}$. For such a potential one can find an analytic solution in terms of parabolic cylinder functions \cite{Birula}.
We calculate the $S$ matrix and the loss probability $P_\mrm{ls}=1-|S_{\ell m}(E)|^2$, assuming QDT boundary conditions \eqref{WKBshort} at the inner side of the barrier, parameterized in terms of $y$ and WKB phase. As we are interested only in the behavior of thermally-averaged reaction rates at large energies where several partial waves contribute, we average $P_\mrm{ls}$ over the WKB phase. This yields
\begin{equation}
\label{Ppar}
P_\mrm{ls}(\ell,y)=\frac{1}{e^{-2\pi \varepsilon(\ell,E)}+1/P^\mrm{re}},
\end{equation}
where $\varepsilon(\ell,E) =E/\hbar\omega_l - \sqrt{2n-4}  \sqrt{\ell(\ell+1)}/(2n)$ is the dimensionless energy measured with respect to the peak of the parabola in units of $\hbar\omega_\ell$. In the universal regime Eq.~\eqref{Ppar} is identical to a textbook solution derived in \cite{Landau}. The reaction rate for a parabolic barrier is given by
\begin{align}
\label{KreSum}
\mathcal{K}^\mrm{re}
\approx
g \frac{\pi \hbar}{\mu k} \sum_{\ell = 0}^{\infty}{^{^\prime} (2\ell+1) \frac{1}{ e^{- 2 \pi \varepsilon(\ell,E)}+1/P^\mrm{re}}}
\end{align}
where $\sum^\prime$ denotes summation over angular momenta allowed by symmetry. At high energies we can replace summation over $\ell$ by integration, which yields
\begin{align}
\label{KreInt}
\mathcal{K}^\mrm{re}  \stackrel{E\to\infty}{\longrightarrow} \frac{\pi \hbar}{\mu k} \int{\frac{d\left[\ell(\ell+1)\right]}{ e^{- 2 \pi \varepsilon(\ell,E)}+1/P^\mrm{re}}}.
\end{align}
The quantum-statistical factor $g$ disappears since for indistinguishable particles the summation is done only over even or odd values of $\ell$.

\begin{figure}
\includegraphics[width=0.9\linewidth]{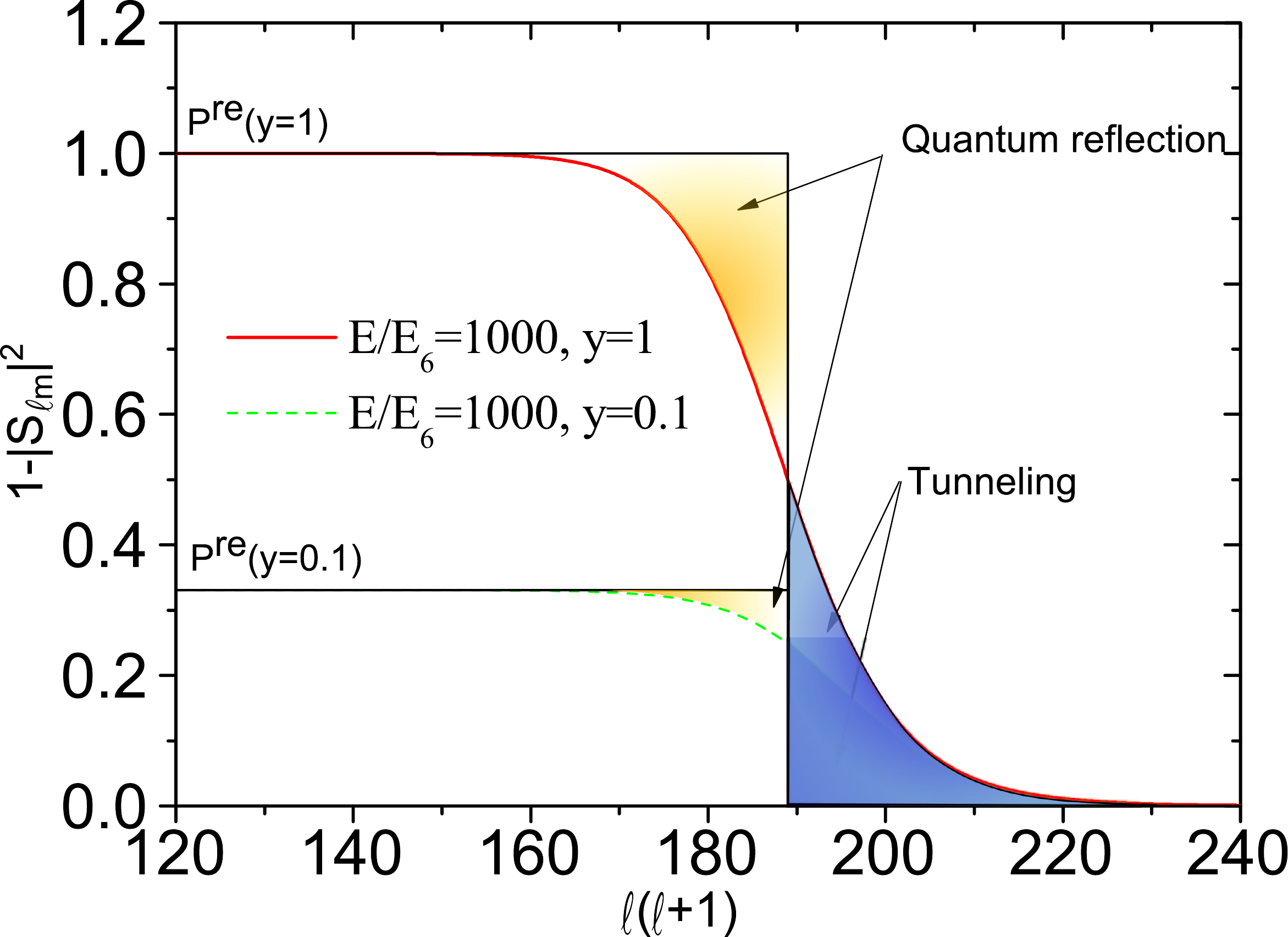}
\caption{
\label{Ferm}
 Loss probability calculated for a parabolic potential fitted to the actual centrifugal barrier of the van der Waals potential and averaged over short-range phase  as a function of a continuous variable $\ell(\ell+1)$ for $y=1$ (left) and $y=0.1$ (right) at energy $E=1000E_6$. 
 The Langevin approximation (black rectangles) assumes constant reaction probability $P^\mrm{re}$ above the barrier, and no reaction below the barrier.}
\end{figure}

The loss probability $P_\mrm{ls}(\ell,y)$ as a function of a continuous variable $\ell(\ell+1)$ is shown in Fig.~\ref{Ferm}. The figure compares the loss probability calculated in the parabolic potential approximation with the classical approach assuming that only collisions with energies above the barrier contribute to the reaction rate. The latter exhibits a step-like behavior, while the former resembles a Fermi-Dirac function. We observe that for $y=1$ the classical description overestimates the reaction rate in the regime affected by the quantum reflection (energies above the barrier, marked in orange), and at the same time does not include the contribution from the quantum tunneling (energies below the barrier, marked in blue). In the universal regime $y=1$ the two contributions turn out to be almost equal, and in this particular case the classical description should work relatively well. In contrast, for $y<1$ the loss probability is additionally affected by the shape resonances which makes the contribution from the quantum tunneling typically larger. In such a case the two effects do not cancel each other and the classical theory underestimates the reaction rate. 

\begin{figure}
\includegraphics[width=0.9\linewidth]{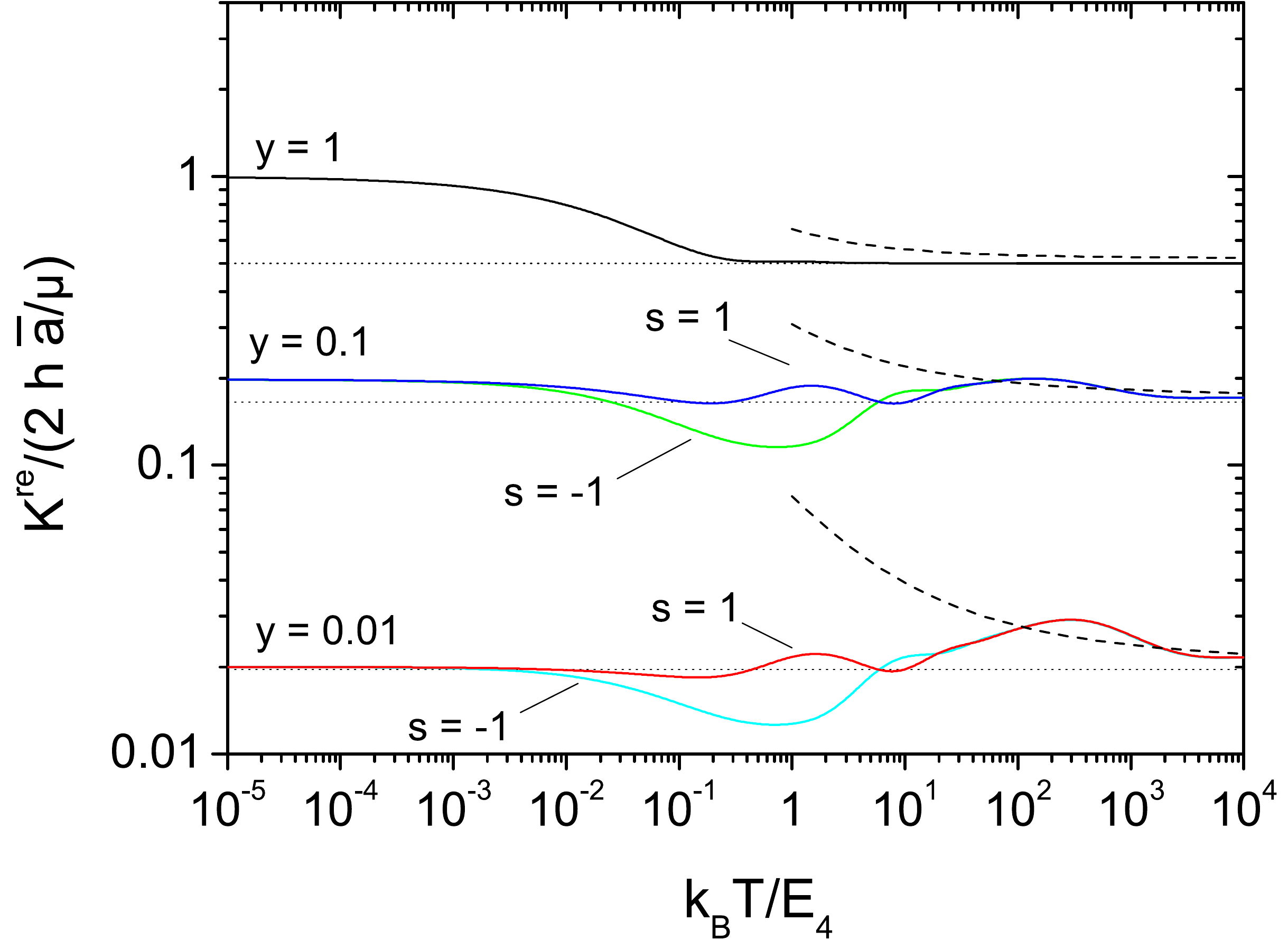}
\caption{
\label{fig:jon}
Thermally averaged reactive rate constants for atom-ion potential, calculated for different reaction amplitudes $y$ and short-range $s$ parameters, along with the Langevin approximation (dotted line) and our high energy treatment (dot-dashed lines).}
\end{figure}

{\it Examples for van der Waals and the polarization potentials.}
\begin{figure}
\includegraphics[width=\linewidth]{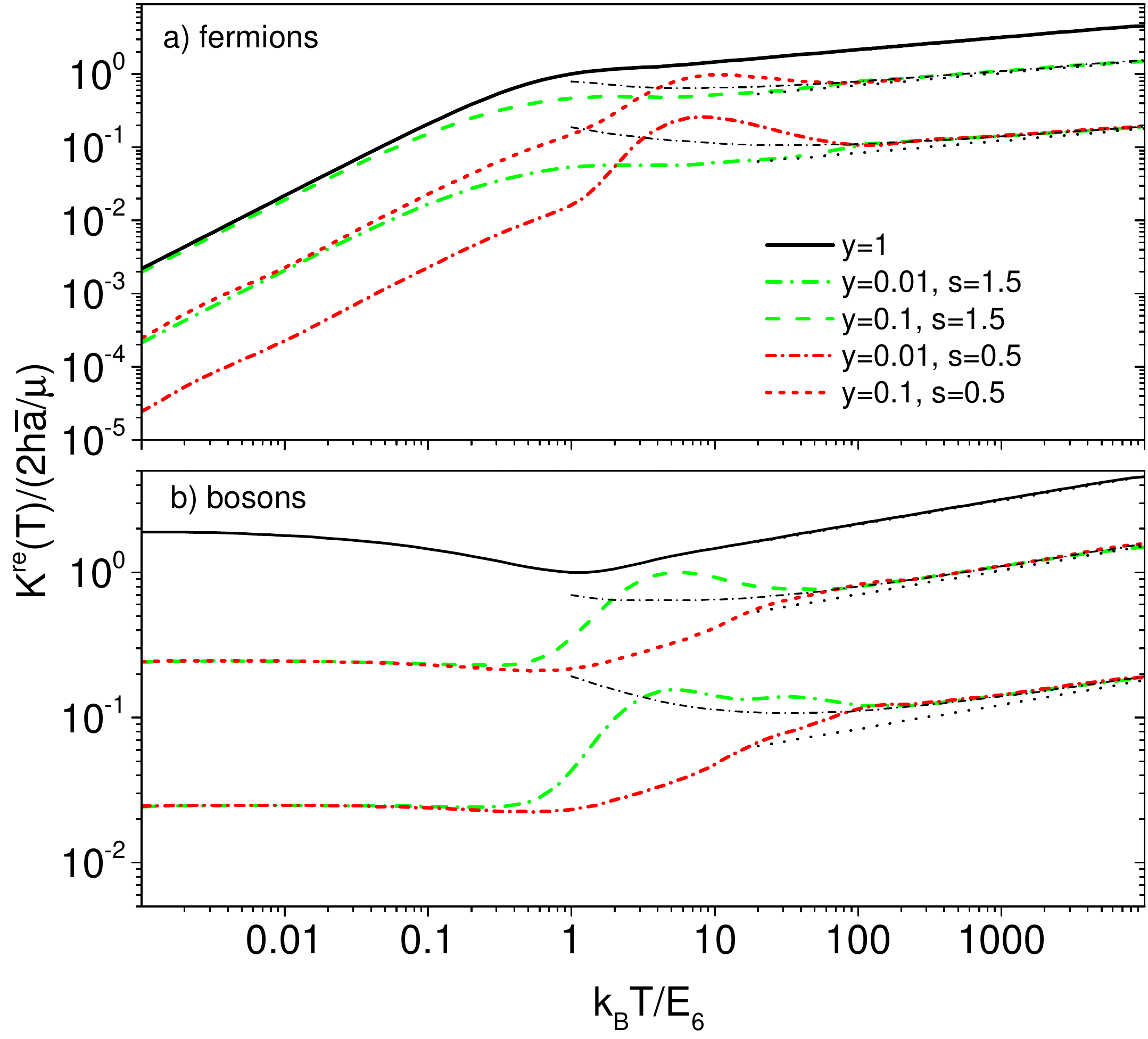}
\caption{
\label{fig:mol}
Thermally averaged reaction rates for van der Waals potential, calculated for different reaction amplitudes $y$ and short-range $s$ parameters, for fermionic and bosonic particles. The dotted lines depict the high-energy approximation \eqref{KLang} assumed in the Langevin theory. The dot-dashed lines show our high-energy approximation \eqref{KreSum} derived by approximating the centrifugal barrier by a parabolic potential.
The universal $y=1$ curve agrees with the result of~\cite{Gao2010PRL}.}
\end{figure}
Fig.~\ref{fig:mol} shows reactive rates for van der Waals potential for bosons and fermions for two exemplary values of $s$ in the universal ($y=1$, considered also in~\cite{Gao2010PRL,Gao2011}) and non-universal ($y=0.1$ and $y=0.01$) regimes. The numerical results (bold solid and dashed lines) were obtained by imposing QDT boundary conditions specified by Eq.~\eqref{WKBshort} at small distances and then propagating the wave function to large distances. After extracting the $S$ matrix we calculate the reactive rates from Eq.~\eqref{Kloss} and then perform thermal averaging: $\langle K_{re}\rangle_{th}(T)=2/\sqrt{\pi}(k_B T)^{(3/2)} \int{dE\,\sqrt{E}e^{-E/k_B T}K_{re}(E)}$.
We note that approximate result \eqref{KreSum} works relatively well in all cases for energies $E \gtrsim 100 E_6$, except for $y=1$ when there is no contribution from the shape resonances. Moreover, we have verified that at large energies the rates for bosons, fermions, and distinguishable particles become equal which agrees with the approximate result of Eq.~\eqref{KreInt}. Figure \ref{fig:jon} shows the reactive rates for distinguishable-particles collisions in the polarization potential, where the high temperature limit is described by the Langevin theory \cite{Langevin1905,Cote2000}. For $n=4$ Eq.~\eqref{KLang} predicts a constant reactive rate. At energies $E \gtrsim 100 E_6$ we find a good agreement of our numerical results with the approximate result \eqref{KreSum} based on the parabolic potential.

\begin{figure}
\centering
\includegraphics[width=0.8\columnwidth]{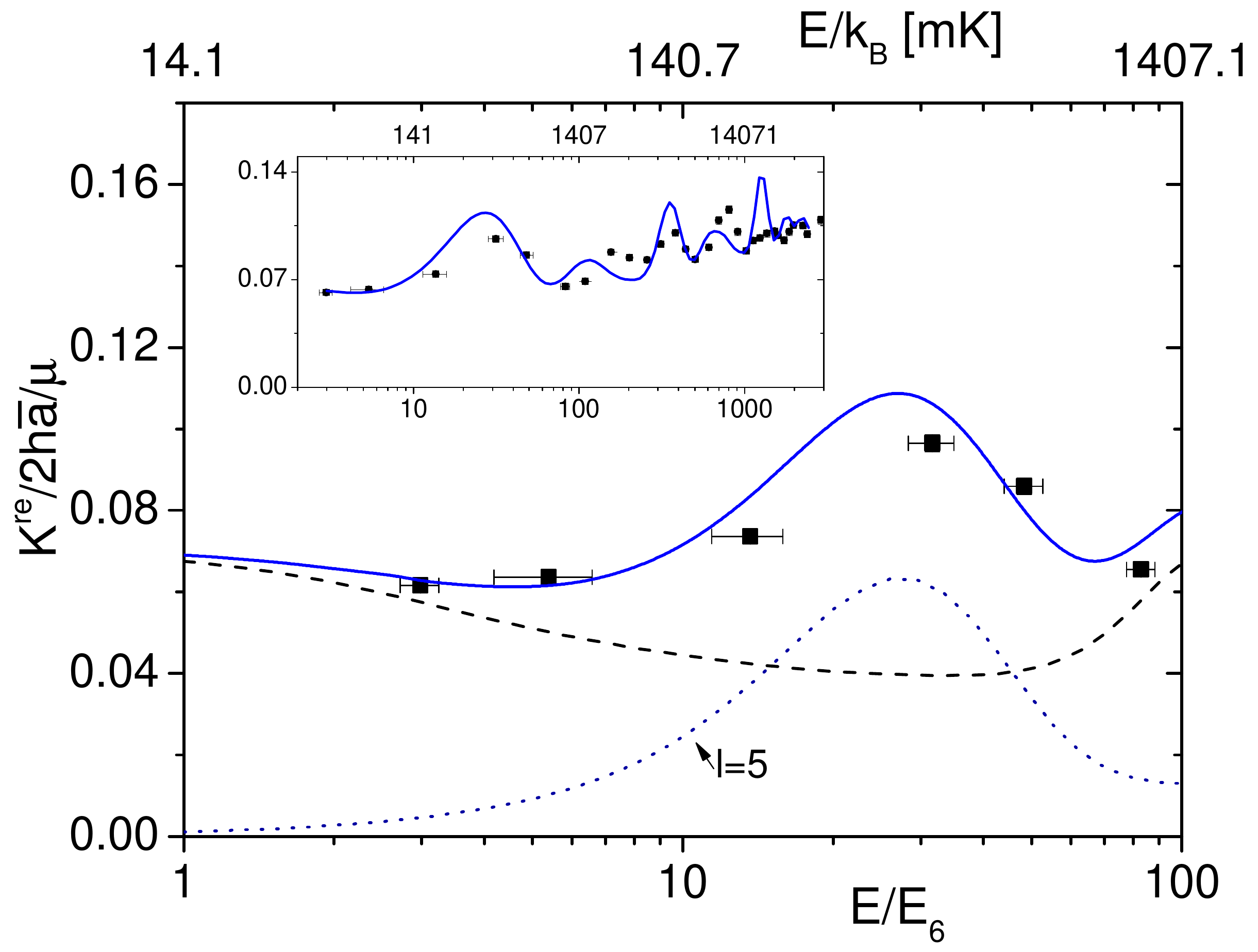}
\caption{
\label{Edv}
Measured reactive rate constants (black points), QDT fit (blue straight line), contribution from $\ell=5$ (dotted line) and other partial waves (dashed line) for the collision of Ar with He$^\ast$. Experimental data is taken from~\cite{Narevicius}. The fit parameters are $y=0.007$, which corresponds to $P^{re}\approx0.028$, and $s=3$. For this system $R_6\approx40\,a_0$ ($a_0$ is the Bohr radius), $E_6/k_B\approx 14$mK and $2h\bar{a}/\mu\approx2.3\cdot10^{-10}$cm$^3 /$s.}
\end{figure}

{\it Penning ionization.}
Finally, we apply our theory to the case of Penning ionization of argon by metastable helium, measured using merged molecular beams in the cold regime~\cite{Narevicius}. The long range part of the Ar-He$^\star$ interaction is given by a van der Waals potential. The short range part can give rise to energy- and $\ell$-dependent corrections to $y$ and $s$. However, we found good agreement with the experimental data by fitting at energies below $\sim100\,E_6$, using the simplest form of our model assuming the same $y$ and $s$ for all partial waves. The position of the resonance is mostly determined by the $s$ parameter, while $y$ sets its width and height. We obtained $y=0.007$ and $s=3$, and the scattering length $a=(58.8-0.7i)a_0$. The experimental data and the the results of scattering calculations with QDT boundary conditions are compared in Fig.~\ref{Edv}. The inset shows the predictions made by extrapolating our fit to the full range of collision energies up to about $30$K. It can be seen that the model gives fairly good agreement with the data. The calculations shown in Fig.~\ref{Edv} were averaged over the velocity distribution of a merged beam that was measured and characterized by the experiment \cite{Narevicius}. We found that the resonance observed near $300$mK is due to the contribution from the single partial wave $\ell=5$.

In conclusion, we present a simple yet general QDT model of a reactive collision between two particles interacting by a power-law potential. We have shown how the low energy collisions are affected by shape resonances and how they contribute to the high energy collisions. Our theory is the first approach to nonuniversal collisions at low energies which does not require calculations of potential energy surfaces and can be applied to various kinds of problems, such as molecular collisions, atom-ion collisions and Penning ionization.

We thank E.~Narevicius for discussions and for providing experimental data. This work was supported by the Foundation for Polish Science International PhD Projects and TEAM programmes co-financed by the EU European Regional Development Fund, National Center for Science grant number DEC-2011/01/B/ST2/02030 and AFOSR MURI Grant FA9550-09-1-0617.
\bibliography{Allrefs}
\end{document}